\documentstyle[12pt]{article}
\def\a{\begin{eqnarray}}
\def\b{\end{eqnarray}}
\def\ax{\begin{equation}\begin{array}{l}}
\def\bx{\end{array}\end{equation}}
\def\0{\nonumber}
\newcommand{\brr}{\begin{array}}
\newcommand{\err}{\end{array}}
\newcommand{\bc}{\begin{center}}
\newcommand{\ec}{\end{center}}

\def\G{{\cal G}}

\def\L{{\Lambda}}
\def\M{{\cal M}}

\def\V{{\cal V}}
\def\l{\lambda}

\def\al{\alpha}
\def\be{\beta}
\def\ep{\epsilon}
\def\o{\omega}

\def\th{\theta}
\newcommand{\W}[5]{W\left(\begin{array}{cc}
\mbox{$#1$}&\mbox{$#2$}\\
\mbox{$#3$}&\mbox{$#4$}\end{array}\bigg|#5\right)}
\newcommand{\Sm}[5]{S\left(\begin{array}{cc}
\mbox{$#1$}&\mbox{$#2$}\\
\mbox{$#3$}&\mbox{$#4$}\end{array}\bigg|#5\right)}

\newcommand{\del}{\partial}
\newcommand{\delb}{\bar\partial}

\begin{document}

\begin{titlepage}

\begin{flushright}
SISSA-ISAS 150/94/FM

\end{flushright}
\vskip0.5cm
\centerline{\LARGE Critical RSOS Models  in External Fields}
\vskip1.5cm
\centerline{\large   I. Vaysburd}
\centerline{International School for Advanced Studies (SISSA/ISAS)}
\centerline{Via Beirut 2, 34014 Trieste, Italy}
\centerline{vaysburd@tsmi19.sissa.it  }
\vskip5cm
\abstract{ We suggest a new family of unitary RSOS scattering models
which is obtained
by placing the $SO(N)$ critical models in "electric" or "magnetic"
field. These fields are associated with two operators from the space
of the $SO(N)$ RCFT corresponding to the highest weight of the vector
representation of $SO(N)$. A perturbation by the external fields
destroys the Weyl group symmetry of an original statistical model.
We show that the resulting kinks scattering theories can be viewed as
affine imaginary Toda models for non-simply-laced and twisted algebras
taken at rational values (roots of unity) of $q$-parameter.
We construct the fundamental kink $S$-matrices for
these models. At the levels $k=1, 2, \infty$ our answers match the
known results for the Sine-Gordon, $Z_{2N}$ - parafermions and free
 fermions respectively.
As a by-product in the $SO(4)$-case
we obtain an RSOS $S$-matrix describing an integrable coupling
of two minimal CFT.   }

\end{titlepage}

\section{Introduction}

The restricted-solid-on-solid models (RSOS)
\cite{abf} - \cite{p} form an important and interesting class of
integrable models. They appear in various mathematical and physical
topics such as statistical mechanics, conformal field theory (CFT),
quantum groups (QG) et.c. A large class of the $2D$ statistical models
(including Ising, $Z_N$ - models and their multicritical versions)
can be reformulated in RSOS terms. It means that one can attach a
finite Lie algebra $\G$ to each of these models. Dynamical variables
$\l$ placed at sites of a $2D$ - lattice belong to a subset of the
integral positive weights of the algebra $\G$ marking the irreducible
representations $\pi_{\l}$. The subset is constrained by

\a
\l \th \leq k
\label {w}
\b
where $\th$ is a highest root of $\G$, $k$ - an integer called a level of
an RSOS.  A remarkable fact is that a critical (multicritical)
behaviour of RSOS is described by some rational $2D$ CFT, namely, by a minimal
models with the W - symmetry \cite {huse}, or coset models

\a
\M_{\G, k} = {{\G_1 \times \G_k}\over
{\G_{k+1}}}
\label {c}
\b

Another crucial observation \cite {az} is that a scaling behaviour
of an RSOS in a vicinity of a critical point is described by some
(1+1) relativistic scattering theory which can be identified with
a massive perturbation of the CFT by a relevant operator from the
Hilbert space. These perturbations usually preserve an infinite
number of charges. It means that the relativistic scattering theory
is a factorized scattering theory (FST) \cite {zz} and it's
S-matrix obeys the natural requirements:
factorizability, crossing symmetry and unitarity.
The RSOS S-matrices are the S-matrices of kinks, interpolating
between different vacua of a massive theory \cite {sm1, bl}.
Remarkably, they are objects of the same nature as local probabilities
of RSOS and can be thought of as some special limit of these probabilities
(generaly taken at another level)
upon imposure of the above requirements.
Exact solutions of RSOS are rarely achievable. Therefore, scaling solutions
given by an S-matrix are very important and can be a subject of a further
investigation by means of TBA \cite {alz}.
The RSOS scattering theories have been intensively studied for the
recent several years \cite {sm1} - \cite{holl}.
D. Gepner suggested a general classification program based on a
fusion ring structure of underlying RCFT.
However, the attention was paid mostly to the certain class of
rational FST (RFST) corresponding to a "thermal" shift from
the critical point
\a
\M^{(t)}_{\G, k} = \M_{\G, k} + g \int  \Phi _{\rho,  \rho + \th}.
\label {th}
\b
Here
\a
\rho = \sum \l_{fund}
\label {rho}
\b
and we are using notations for RCFT primaries introduced in \cite {fl}.
Thermal deformation preserves a discrete group of original RSOS
(the Weil group of $\G$). We briefly review these models in Sec.2

The exceptions are the $\Phi _{1,2}$ and $\Phi_{2,1}$ perturbations
of minimal CFT destroying $Z_2$ symmetry
considered by F. Smirnov \cite {sm2} and the perturbations of
critical parafermions by the first parafermionic current destroying
 $Z_N$ (V. Fateev \cite {fat}).
In both cases the interesting kink systems are constructed.
In Sec.3 we present a large class of RFST  including as the
special cases the examples of
F. Smirnov and V. Fateev.
We show that there are two integrable primaries in each of the considered
RCFT which violate the Weyl group of $\G$ in a way similar to
the $Z_2$ - violation by the magnetic field coupled to the
spin density operator of the critical Ising model. In order
to stress this analogy we call the two perturbations "
electric" and "magnetic" and the resulting RCFT -
$\M^{(e)}_{\G, k}$ and $\M^{(h)}_{\G, k}$ respectively
\a
\M^{(e,h)}_{\G, k} = \M_{\G, k} + g \int  \Phi ^{(e,h)}
\label {eh}
\b
\a
\Phi^{(h)} \equiv \Phi_{ \rho, \rho + \o} \0\\
\\
\Phi^{(e)} \equiv \Phi_{ \rho + \o, \rho} \0\\
\b
We denote by $\o$ the highest weight of the vector representation
of $\G$.

In Sec. 4 we investigate a quantum symmetry of the constructed
RFST and argue that there is a twisted affine algebra attached
to each of them according to some rule. It worthwhile mentioning
that thermal models can be viewed as restricted complex ATFT
obtained by the highest root affinization $\G \rightarrow \G^{(1)}$
as long as $\M^{(e,h)}$ correspond to non-simply-laced complex
ATFT taken at rational points of a q-deformation parameter.
A quantization of the complex ATFT for non-simply-laced algebras
is a challenging problem \cite {holl}.
For the real coupling constant the solution found in \cite {dgz}
and studied in \cite {cds} exhibits an interesting duality property.
In a sense the S-matrix solution for $\G = SO(N)$
suggested in Sec. 5 is a solution
for the complex non-simply-laced ATFT
at "unitary" points and we hope that it will
shed light on the general situation.

In Sec. 6 we compare the solution to the known answers at the points
where our family of models intersects with some families of
$\M^{(t)}$ - type.

Sec. 7 is devoted to discussions.

\section{Thermal perturbations of minimal W-models and
RSOS - scattering}

Consider an RSOS on a $2D$ lattice whose critical regime is described
by $\M_{\G, k}$. For each link $(ij)$ of the lattice the variables
$\l_i$ and $\l_j$ obey the admissibility condition expressed by the following
relations
\ax
\pi_{\l_j}\in \pi_{\l_i}\otimes \pi_1\0\\
\pi_{\l_i}\in \pi_{\l_j}\otimes \pi_{\bar{1}}
\bx
$\phi_1$ and $\pi_{\bar 1}$ stand for fundamental and antifundamental
representations of $\G$. If we denote by dots possible values of $\l$
selected by (\ref{w}) and connect by a link each pair of admissible
$\l$'s then we obtain a graph  $G$ with a finite number of nodes.
RSOS configurations are given by different embeddings of the lattice
into this graph. Local probabilities $\W{a}{b}{c}{d}{u}$
attached to each plaquette
of the lattice are functions of an embedding of this plaquett and a
spectral parameter $u$. $a,b,c,d$ mark vertices of $G$.
An integrability condition
of the RSOS is expressed
by the star - triangle equation on the local probabilities
\a
\sum_b\W{a}{b}{c}{d}{u}\W{e}{f}{b}{d}{v}\W{g}{e}{a}{b}{u+v}\0\\
=\sum_b \W{g}{e}{b}{f}{u}\W{g}{b}{a}{c}{v}\W{b}{f}{c}{d}{u+v}
\label{ste}
\b
Solutions of (\ref{ste}) are found for the $A$,$B$,$C$ and $D$
algebras \cite{jmo}. In the last three cases it was implied
that the graph G is generated by $\pi_{\o}$.
A remarkable property of these solutions is their rotational
symmetry
\a
\W{a}{b}{d}{c}{u} = \left({\phi_a \phi_c \over \phi_b \phi_d}\right)^{-1/2}
\W{d}{a}{c}{b}{\L - u},
\label{rs}
\b
where $\phi_{\l}$ stands for the quantum dimention of the representation
$\pi_{\l}$ at the level $k$. $\L$ is a crossing parameter depending on
$k$ and the Coxeter number $c_{\G}$. The rotational symmetry is closely
related  to the crossing symmetry in $2D$ scattering. Another property
connected with FST is quasiunitarity
\a
\sum_d \W{a}{b}{c}{d}{u} \W{d}{b}{c}{e}{-u} = f(u) f(-u) \delta _{ae}
\label{qu}
\b
For the $B$,$C$ and $D$ algebras
\a
f(u) = {\sin (\o -u)\;\; \sin (\L -u) \over \sin \o \;\; \sin \L}
\label{fu}
\b
\a
\o={i\pi \over c_{\G} + k}\\
\\
\L = \o c_{\G}  /2 \0
\b
The solutions of \cite {jmo} form a family parametrized by
the elliptic parameter $P$. In the limit $P\to\infty$
local probabilities become simple combinations of trigonometric
functions and the system exhibits a critical behaviour.
Small deviation from zero in the $P$ - direction is described
by the thermal perturbation (\ref{th}) of a RCFT. The corresponding
$2D$ field theory possess an infinite number of IM. Their spins
measured by the operator $s=L_0 - \bar L_0$ from $2D$
conformal algebra follow Coxeter exponents of $\G$ modulo $c_{\G}$
with some exceptions for low $k$. An $S$-matrix of kinks
which interpolate between the vacua of the theory satisfies (\ref {ste})
with exchange $k\to k-1$.
This $S$-matrix can be obtained from the local probabilities $W$
by the well known procedure. It was done first in \cite {sm1,
smresh, bl} for $\G = A_1$. The answer is the restricted Sine - Gordon
$S$-matrix. V. Fateev and H. de Vega generalized this result for
$\G = A_n$, $n\geq 2$. The work for the rest of the algebras has
been done by D. Gepner \cite {gep1, gep2}. The $C$ - algebras have
been treated in \cite {holl}.
An $S$-matrix element corresponding to a kink-kink scattering
process
\a
K_{ab} + K_{bd} \to K_{ac} + K_{cd}
\label{kk}
\b
is given by
\a
\Sm{a}{b}{c}{d}{\th} = F (\th\L/i\pi)
\left({\phi_a \phi_d \over \phi_b \phi_c}\right)^
{\th\over 2\pi i}
 \W{a}{b}{c}{d}{\th\L/i\pi}
\label {sw}
\b
$\th$ is a relative rapidity of the in-coming and out-coming kinks.
And $F(u)$ is a minimal solution of the equations
\a
F(u) = F(\L-u)
\label {F}
\b
\a
{F(u)F(-u)\over f(u)f(-u)} = 1
\label {uF}
\b
The $S$-matrix constructed in such a way satisfies

\noindent
{\bf crossing}
\a
\Sm{a}{b}{c}{d}{\th} = \Sm{c}{a}{d}{b}{i\pi - \th}
\label {cross}
\b

\noindent
and
{\bf unitarity}
\a
\sum_d \Sm{a}{b}{c}{d}{\th} \Sm{d}{b}{c}{e}{-\th} =  \delta _{ae}
\label{unit}
\b
The star-triangle equation is not violated by additional factors
in (\ref{sw}).

\section{Integrability of the  vector perturbations of the $SO(N)$ RCFT}

The thermal shifts along the $\Phi_{\rho, \rho + \th}$ - direction
do not exhaust all the integrable directions in a parametric space
near the criticallity.
It turns out that for a sufficiently large class of RCFT
one can point out other integrable deformations
\cite{vays} corresponding to "external fields" (\ref{eh}).
For $\G = SU(2), SU(3)$ and $SO(N)$ their anomalous
dimensions are

SU(2):
\ax
\Delta^{1,2}_p = {1 \over 4} (1 - {3 \over p + 1})
\label{del0}\\
\Delta^{2,1}_p = {1 \over 4} (1 + {3 \over p })
\label{del1}
\bx

SU(3):
\ax
\Delta^{(h)}_p = {1 \over 3} (1 - {4 \over p + 1})
\label{del3}\\
\Delta^{(e)}_p = {1 \over 3} (1 + {4 \over p })
\label{del2}
\bx

SO(N):
\ax
\Delta^{(h)}_p = {1 \over 2} (1 - {N-1 \over p+1})\\
\Delta^{(e)}_p = {1 \over 2} (1 + {N-1\over p })
\label{del}
\bx
\a
p\equiv c_{\G} + k\0
\b

The integrability of the resulting massive models defined by (\ref{eh})
can be proven by the counting argument .
To this purpose let us note that both operators - $\Phi^{(h)}$ and $\Phi^{(e)}$
 - are the most relevant ones in the operator algebras generated by each of
them.
So, in order to establish integrability one needs just to compare
multiplicities
appearing in the expansions of the conformal characters
$\V_0/\del \V_0$ and $\V^{(e,h)}/\del \V^{(e,h)}$
\a
\chi_{\V_0/\del \V_0}(q) = (1-q)\chi_{\V_0}(q) + q = \sum_s a_s q^s,
\0
\b
\a
\chi_{\V^{(e,h)}/\del \V^{(e,h)}}(q) = (1-q) \chi_{\V^{(e,h)}}(q)
= \sum_s b_s^{(e,h)} q^s \0
\b

For the characters of the operators (\ref{del0} -
\ref{del2})
one has
\a
a_6 = b_5 + 1
\b
for (\ref{del})
\a
a_4 = b_3^{(e,h)} + 1
\b
This means the existence of the nontrivial IM
of spin $s=5$ and $s=3$ respectively.
This fact for $\Phi_{1,2}$ and
$\Phi_{2,1}$
operators in unitary minimal models has been known long ego
\cite{az}. In a sense we consider the perturbations of RCFT
generalizing the latter ones treated in \cite{sm2}.
The models $\M^{(e,h)}_{SO(4), k}$ are somewhat curious.
In this case original CFT is nothing but a tensor product
of two copies of a minimal model because $SO(4)=SU(2)\times
SU(2)$. The perturbations which couple the two copies are
\a
\Phi^{(h)} = \Phi^{(1)}_{1,2}\;\Phi^{(2)}_{1,2}\\
\Phi^{(e)} = \Phi^{(1)}_{2,1}\;\Phi^{(2)}_{2,1}
\label{so}
\b
thus
\a
\M_{SO(4), k}^{(e,h)} = \M_k^{(1)} + \M_k^{(2)} + g\int \Phi^{(e,h)}
\label{so1}
\b
Both of these models enjoy $Z_2$ symmetry permuting the first and the second
copy. The conserved charge of spin $s=3$ can be constructed explicitely
starting from two holomorphic currents of spin $s=4$ respecting this
permutation
\a
J_1(z) = :\Big(T^{(1)}\Big)^2: + :\Big(T^{(2)}\Big)^2:
\label{j1}
\b
\a
J_2(z) = :T^{(1)} T^{(2)}:
\label{j2}
\b
These two currents are opposed by the only $s=3$ non-derivative
descendant of $\Phi^{(1)}\;\Phi^{(2)}$:
$$L_{-3}\Phi^{(1)}\;\Phi^{(2)} + \Phi^{(1)}\;L_{-3}\Phi^{(2)}.$$
This descendant drops out of the r.h.s. of conservation law of the
following linear combination:
$$J(z) = J_1(z) + {2(\Delta-1)\over \Delta} J_2(z),$$
hence
\ax
\delb J = g\del \L_2 (z, \bar z),
\bx
where $\Delta = \Delta^{(e,h)}_{4,p}$ is given by
(\ref{del}).

Integrability of the $\M^{(h)}_{3,p}$ has been discovered in \cite{matdep}.
Examination of W - characters allows to conjecture
that in general the models (\ref{del1},\ref{del2}) exhibit IM's at spins
\a
s=1,5,7,11, \cdots
\0
\b
and the models (\ref {del}) at spins
\a
s=1,3,5,7,\cdots
\0
\b
for sufficiently large $p$'s.
\section{Quantum symmetry}

In order to find an RSOS S-matrix for the integrable models
constructed in Sec.3 one has to choose between the Yang - Baxter
solutions to start from. These solutions are naturally
marked by the affine Lie algebras \cite{jmo}.
So, we have to understand, which of the affine algebras
correspond to our models.
The question is resolved by the following

{\bf Theorem}

\noindent
{\em Consider a RCFT given by (\ref{c}) and an integrable massive
perturbation by a primary $\Phi_{\rho,\rho + \l}$
(or $\Phi_{\rho + \l,\rho}$ ). The resulting RFST corresponds
to a $q$-deformed affine Lie algebra whose Dynkin diagram
 $D(\G,\l)$ can be obtained by attaching additional $-\l$
root to the finite diagram $D(\G)$ and ${\bf inversion}$ of the
arrow (if any) connecting this root with the finite part
(fig.\ref{fig1})}


\begin{figure}
\centering
\begin{picture}(200,200)(- 100,- 170)

\thinlines
\put(-80, 0){\line(-1,1){7}}
\put(-80, 0){\line(-1,-1){7}}
\put(-80, 0){\line(1,0){10}}
\put(-70, 0){\line(1,0){5}}
\put(-50, 0){\line(-1,0){5}}
\put(-50, 0){\line(1,0){10}}
\put(-40, 0){\line(1,-1){7}}

\put(40, 0){\line(-1,1){7}}
\put(40, 0){\line(-1,-1){7}}
\put(40, 0){\line(1,0){10}}
\put(50, 0){\line(1,0){5}}
\put(70, 0){\line(-1,0){5}}
\put(70, 0){\line(1,0){10}}
\put(80, 0){\line(1,1){7}}
\put(80, 0){\line(1,-1){7}}

\put(-60,-60){\makebox(0,0){${\bf >}$}}
\put(60,-60){\makebox(0,0){${\bf <}$}}
\put(-35,-90){\makebox(0,0){${\bf >}$}}
\put(85,-90){\makebox(0,0){${\bf <}$}}
\put(-35,-120){\makebox(0,0){${\bf >}$}}
\put(85,-120){\makebox(0,0){${\bf <}$}}
\put(-52.5,-150){\makebox(0,0){${\bf >}$}}
\put(67.5,-150){\makebox(0,0){${\bf <}$}}
\put(-60,0){\makebox(0,0){$\dots$}}
\put(60,0){\makebox(0,0){$\dots$}}
\put(-60,10){\makebox(0,0){$D_n$}}
\put(60,10){\makebox(0,0){$D_n^{(1)}$}}
\put(-27,7){\makebox(0,0){$-\theta$}}
\put(-105,-15){\makebox(0,0){$a)$}}

\put(-80,0){\circle*{2}}
\put(-70,0){\circle*{2}}
\put(-50,0){\circle*{2}}
\put(-40,0){\circle*{2}}
\put(40,0){\circle*{2}}
\put(50,0){\circle*{2}}
\put(70,0){\circle*{2}}
\put(80,0){\circle*{2}}

\put(-87.1,7.1){\circle*{2}}
\put(-87.1,-7.1){\circle*{2}}
\put(87.1,7.1){\circle*{2}}
\put(87.1,-7.1){\circle*{2}}
\put(-32.9,7.1){\circle*{2}}
\put(-32.9,-7.1){\circle*{2}}
\put(32.9,7.1){\circle*{2}}
\put(32.9,-7.1){\circle*{2}}

\thicklines
\put(-40, 0){\line(1,1){7}}
\put(- 15,0){\vector(1,0){30}}
\thinlines
\put(-80, -29){\line(-1,0){10}}
\put(-80, -31){\line(-1,0){10}}
\put(-80, -30){\line(1,0){10}}
\put(-70, -30){\line(1,0){5}}
\put(-50, -30){\line(-1,0){5}}
\put(-50, -30){\line(1,0){10}}
\put(-40, -30){\line(1,-1){7}}
\thicklines
\put(-84, -28){\line(-1,-1){2}}
\put(-84, -32){\line(-1, 1){2}}
\put(36, -28){\line(-1,-1){2}}
\put(36, -32){\line(-1, 1){2}}
\thinlines
\put(40, -29){\line(-1,0){10}}
\put(40, -31){\line(-1,0){10}}
\put(40, -30){\line(1,0){10}}
\put(50, -30){\line(1,0){5}}
\put(70, -30){\line(-1,0){5}}
\put(70, -30){\line(1,0){10}}
\put(80, -30){\line(1,1){7}}
\put(80, -30){\line(1,-1){7}}

\put(-60,-30){\makebox(0,0){$\dots$}}
\put(60,-30){\makebox(0,0){$\dots$}}
\put(-60,-20){\makebox(0,0){$B_n$}}
\put(60,-20){\makebox(0,0){$B_n^{(1)}$}}
\put(-27,-23){\makebox(0,0){$-\theta$}}

\put(-80,-30){\circle*{2}}
\put(-70,-30){\circle*{2}}
\put(-50,-30){\circle*{2}}
\put(-40,-30){\circle*{2}}
\put(40,-30){\circle*{2}}
\put(50,-30){\circle*{2}}
\put(70,-30){\circle*{2}}
\put(80,-30){\circle*{2}}

\put(-90,-30){\circle*{2}}
\put(87.1,-22.9){\circle*{2}}
\put(87.1,-37.1){\circle*{2}}
\put(-32.9,-22.9){\circle*{2}}
\put(-32.9,-37.1){\circle*{2}}
\put(30,-30){\circle*{2}}

\thicklines
\put(-40, -30){\line(1,1){7}}
\put(- 15,-30){\vector(1,0){30}}

\thinlines
\put(-70,-59){\line(1,0){20}}
\put(-70,-60){\line(1,0){20}}
\put(-70,-61){\line(1,0){20}}
\put(-70,-62){\line(1,0){20}}
\put(50,-59){\line(1,0){20}}
\put(50,-60){\line(1,0){20}}
\put(50,-61){\line(1,0){20}}
\put(50,-62){\line(1,0){20}}
\put(-62,-58){\line(1,-1){2.5}}
\put(-62,-63){\line(1,1){2.5}}
\put(62,-58){\line(-1,-1){2.5}}
\put(62,-63){\line(-1,1){2.5}}
\thicklines
\put(- 15,-60){\vector(1,0){30}}

\put(-70,-60.5){\circle*{3}}
\put(-50,-60.5){\circle*{3}}
\put(50,-60.5){\circle*{3}}
\put(70,-60.5){\circle*{3}}

\put(-70,-55){\makebox(0,0){$A_1$}}
\put(60,-50){\makebox(0,0){$A_2^{(2)}$}}
\put(-52,-55){\makebox(0,0){$-\omega$}}
\put(-105,-60){\makebox(0,0){$b)$}}
\thinlines
\put(-80, -90){\line(-1,1){7}}
\put(-80, -90){\line(-1,-1){7}}
\put(-80, -90){\line(1,0){10}}
\put(-70, -90){\line(1,0){5}}
\put(-50, -90){\line(-1,0){5}}
\put(-50, -90){\line(1,0){10}}
\put(-40, -91){\line(1,0){10}}
\put(-40, -89){\line(1,0){10}}

\put(40, -90){\line(-1,1){7}}
\put(40, -90){\line(-1,-1){7}}
\put(40, -90){\line(1,0){10}}
\put(50, -90){\line(1,0){5}}
\put(70, -90){\line(-1,0){5}}
\put(70, -90){\line(1,0){10}}
\put(80, -91){\line(1,0){10}}
\put(80, -89){\line(1,0){10}}

\put(-60,-90){\makebox(0,0){$\dots$}}
\put(60,-90){\makebox(0,0){$\dots$}}
\put(-60,-80){\makebox(0,0){$D_n$}}
\put(60,-80){\makebox(0,0){$A_{2n-1}^{(2)}$}}
\put(-27,-85){\makebox(0,0){$-\omega$}}
\put(-105,-105){\makebox(0,0){$c)$}}
\put(-105,-150){\makebox(0,0){$d)$}}

\put(-80,-90){\circle*{2}}
\put(-70,-90){\circle*{2}}
\put(-50,-90){\circle*{2}}
\put(-40,-90){\circle*{2}}
\put(40,-90){\circle*{2}}
\put(50,-90){\circle*{2}}
\put(70,-90){\circle*{2}}
\put(80,-90){\circle*{2}}

\put(-87.1,-82.9){\circle*{2}}
\put(-87.1,-97.1){\circle*{2}}
\put(90,-90){\circle*{2}}
\put(32.9,-82.9){\circle*{2}}
\put(32.9,-97.1){\circle*{2}}
\put(-30,-90){\circle*{2}}

\thicklines
\put(- 15,-90){\vector(1,0){30}}
\put(- 15,-150){\vector(1,0){30}}
\thinlines
\put(-80, -119){\line(-1,0){10}}
\put(-80, -121){\line(-1,0){10}}
\put(-80, -120){\line(1,0){10}}
\put(-70, -120){\line(1,0){5}}
\put(-50, -120){\line(-1,0){5}}
\put(-50, -120){\line(1,0){10}}
\put(-40, -30){\line(1,-1){7}}

\put(-84, -118){\line(-1,-1){2}}
\put(-84, -122){\line(-1, 1){2}}
\put(36, -118){\line(-1,-1){2}}
\put(36, -122){\line(-1, 1){2}}
\thinlines
\put(40, -119){\line(-1,0){10}}
\put(40, -121){\line(-1,0){10}}
\put(40, -120){\line(1,0){10}}
\put(50, -120){\line(1,0){5}}
\put(70, -120){\line(-1,0){5}}
\put(70, -120){\line(1,0){10}}
\put(80, -121){\line(1,0){10}}
\put(80, -119){\line(1,0){10}}
\put(-40, -121){\line(1,0){10}}
\put(-40, -119){\line(1,0){10}}

\put(-60,-120){\makebox(0,0){$\dots$}}
\put(60,-120){\makebox(0,0){$\dots$}}
\put(-60,-110){\makebox(0,0){$B_n$}}
\put(60,-110){\makebox(0,0){$A_{2n}^{(2)}$}}
\put(-30,-115){\makebox(0,0){$-\omega$}}

\put(-80,-120){\circle*{2}}
\put(-70,-120){\circle*{2}}
\put(-50,-120){\circle*{2}}
\put(-40,-120){\circle*{2}}
\put(40,-120){\circle*{2}}
\put(50,-120){\circle*{2}}
\put(70,-120){\circle*{2}}
\put(80,-120){\circle*{2}}

\put(-90,-120){\circle*{2}}
\put(90,-120){\circle*{2}}
\put(-30,-120){\circle*{2}}
\put(30,-120){\circle*{2}}

\thicklines
\put(- 15,-120){\vector(1,0){30}}
\thinlines
\put(-60,-150){\line(-1,0){15}}
\put(60,-150){\line(-1,0){15}}
\put(-60,-149){\line(1,0){15}}
\put(-60,-150){\line(1,0){15}}
\put(-60,-151){\line(1,0){15}}
\put(60,-149){\line(1,0){15}}
\put(60,-150){\line(1,0){15}}
\put(60,-151){\line(1,0){15}}
\put(-75,-150){\circle*{2}}
\put(-60,-150){\circle*{2}}
\put(-45,-150){\circle*{2}}
\put(75,-150){\circle*{2}}
\put(60,-150){\circle*{2}}
\put(45,-150){\circle*{2}}

\put(-67.5,-145){\makebox(0,0){$A_2$}}
\put(60,-140){\makebox(0,0){$D_4^{(3)}$}}
\put(-43,-145){\makebox(0,0){$-\omega$}}

\end{picture}

\caption[x]{\footnotesize

a) Thermal perturbation. No arrow inversion.

b) $A^{(2)}_2$ - case. Inversion does not change the diagram.

c)$(e,h)$ - perturbations of $SO(N)$ theories.

d)$(e,h)$ - perturbations of $SU(3)$ theory.}
\label{fig1}
\end{figure}

It should be noted that the theorem is closely related to
the duality between a root system of a Toda lattice and
a set of the nonlocal charges observed in \cite{flec}.
We will skip a rigorous proof of this theorem presenting
just several checks for it which are quite convincing however.
The simplest check is to examine a symmetry of a model in the
rational limit $p\to \infty$.
In such a limit the  nonlocal charges \cite{bl2}
become local IM and can be easily constructed.

\noindent
$${\bf \forall\G ; \;\;\;\l = \th}$$

\noindent
This case corresponds to
thermal RSOS. $$ D(\G,\th) = D(\G^{(1)}) $$ Therefore we have
to deal with $\G^{(1)}$ S-matrices in agreement with \cite
{sm1,fdv,gep1,gep2}.
 \vskip 0.2in
\noindent
 $${\bf \G = A_1; \;\; \l = \o}$$

\noindent
For this case we have
$$ D(A_1,\o) = D(A_2^{(2)}).$$ The inversion does not change
the diagram (fig.1) So, the S-matrix solution of the model can
be obtained from the $A_2^{(2)}$ solution of YB equation
as it was done in \cite {sm2, ik}. IM's generating $A_2^{(2)}$
in the rational limit are constructed in \cite{vays}.
The nonlocal charges for $p<\infty$
\cite{costas} form the $A_{2,q}^{(2)}$
algebra at $q^p=1$.

\noindent
 $${\bf \G = A_2; \;\; \l = \o}$$

\noindent
In this case (fig. 1)
$$ D(A_2,\o) = D(D_4^{(3)}).$$
In the rational limit the model coincides with the complex
$A_2^{(1)}$ Toda model at the second reflectionless point $\Delta_{pert}
=1/3$
\cite{vays}. One can explicitly construct $D_4^{(3)}$
Noether charges acting on the 8-plet of the lightest particles:
six solitons and antisolitons plus two breathers.
(see \cite{vays} for details). The 8-plet transforms in the
vector representation of the $D_4^{(3)}$.

\noindent
$${\bf \G = SO(N);\;\; \l = \o}$$

\noindent
This is the case to which the main
attention is going to be paid in the rest of the paper.
$$ D (SO(N), \o) = D(A_{N-1}^{(2)}).$$
Once again it is very instructive to look at the rational limit of the
model. We immediately see that in this limit the model is nothing else
but $N$ free massive fermions. As was observed in \cite {sotkov, vays}
such a system exhibits the charges generating twisted affine algebra
$A_{N-1}^{(2)}.$

So,  the theorem  agrees with the known cases.
But a decisive support for it should be provided by the S-matrix
construction of the next Section. This construction is done for the  case
$(SO(N), \;\o)$.
It is
 using the
theorem as an input and shows up a perfect agreement with known
results at all the checkpoints.

\section{ Fundamental kink-kink S-matrices}

First, we construct an admissibility graph $G$
describing the kink-kink scattering in the models (\ref{eh})
for $\G=SO(N)$. We will call them $\M^{(e,h)}_{N,p}$.
The structure of this graph follows from the fusion algebra
of original CFT. It is important to mention that
\a
\Phi^{(e)}(z) \; \Phi^{(h)}(w) = {\Phi_{\rho + \o,\rho + \o}\over
z - w}
\label{fusion}
\b
It means that in the $\M^{(e)}_{N,p}$ model an UV limit of the kink
is given by the operator $\Phi^{(h)}$. The vacua of the theory correspond
to the primaries from the fusion ring generated by $\Phi^{(h)}$. Two vacua
are connected by a link if one of them can be obtained by the fusion of
 another one with the generating operator. The graph constructed in such a
way describes also a finite ring structure of  the representations of the
group $SO(N)_q$ at
\a
q=e^{-i\pi/p+1}
\label{qe}
\b
generated by tensoring of $\pi_{\o}$.

In the  $\M^{(h)}_{N,p}$ model an UV limit of the kink is given
by the operator $\Phi^{(e)}$ which defines a graph $G$. In this
case it corresponds to a representation ring of $SO(N)_q$ at
\a
q=e^{i\pi/p}
\label{qh}
\b
As an example in fig.2 we have drawn the admissibility graphs
at levels
$k=1,2$.
It is important to note  that fundamental
kinks live in the {\em vector}
representation of the quantum group in contrast with the thermal
RSOS when kinks belong to {\em fundamental} (spinorial) representations.
This is the main kinematic difference between $\M^{(e,h)}$ and
 $\M^{(t)}$. It can be viewed as two different ways to affinize
$SO(N)$ algebras - the first one giving $A_{N-1}^{(2)}$ and the
second $B^{(1)}$, $D^{(1)}$ respectively.
So, the local probabilities of \cite{jmo} for $(SO(N),\o)$ in the trigometric
limit seem to be
a suitable input for an $S^{(e,h)}$-matrix
construction. Of course, some changes
are necessary. Otherwise, we will get just a subsector of a thermal
theory with a spinor-antispinor sector missed \cite{gep1}.
It is natural to assume that the appropriate change of the crossing
 parameter $\L $ in the RSOS weights
should be similar to the one in a vertex R-matrix case
 \cite{jimbo}. The right choice of the crossing parameter is given by
\a
\L_{N,p}^{(e,h)} = 1/2(N\o^{(e,h)} + i\pi),
\label{crossing}
\b
where
\a
\o^{(h)}=i\pi/p\;\;\;\;\\
\o^{(e)}=-i\pi/p+1\label{omega}
\b
The main ingredient of the S-matrix (\ref{sw})-
local probabilities $W$ - should be borrowed from \cite{jmo}:
\a
\W{a}{a+\mu}{a+\mu}{a+2\mu}{u}={[\L - u][\o - u]\over [\L][\o]} (\mu\neq 0),
\b
\a
\W{a}{a+\mu}{a+\mu}{a+\nu+\mu}{u}={[\L - u][a_{\mu} -
a_{\nu} + u]\over [\L][a_{\mu} -
a_{\nu}]} (\mu\neq \pm\nu),
\b
\a
\W{a}{a+\nu}{a+\mu}{a+\mu+\nu}{u}={[\L - u][u]\over [\L][\o]}
\left({[a_{\mu} - a_{\nu} + \o][a_{\mu} - a_{\nu} - \o]
\over [a_{\mu} - a_{\nu}]^2}\right)^{1/2} (\mu\neq \pm\nu),
\b
\a
\W{a}{a+\nu}{a+\mu}{a}{u}={[a_{\mu}+a_{\nu}+\o -\L + u][u]\over
[\L][a_{\mu}+a_{\nu}+\o]} {\phi_a\over(\phi_{a+\mu}\phi_{a+\nu})
^{1/2}}\;\;(\mu\neq \nu),
\b
\a
\W{a}{a+\mu}{a+\mu}{a}{u}={[\L + u][2a_{\mu}+\o+2\L - u]
\over [\L][2a_{\mu}+\o+2\L]} - {[u][2a_{\mu}+\o+\L - u]
\over [\L][2a_{\mu}+\o+2\L]} H_{a\mu},\\
={[\L + u][2a_{\mu}+\o + u]
\over [\L][2a_{\mu}+\o]} - {[u][2a_{\mu}+\o-\L+ u]
\over [\L][2a_{\mu}+\o]} {\phi_a\over\phi_{a+\mu}}\;(\mu\neq 0),
\b
Here $[u]\equiv\sinh u$; $\mu,\nu$ are weights from $\pi_{\o}$,
$$a_{\mu}=[(a+\rho,\mu) -{1\over 2}\delta_{0,\mu}]\o$$ and
$$H_{a\mu}=\sum_{\nu\neq\mu}{[a_{\nu}+a_{\mu}+\o+2\L]\over
[a_{\nu}+a_{\mu}+\o]}{\phi_a\over\phi_{a+\nu}}.$$
In the last formula it is implied that $a$ is admissible
with $a+\nu$.

Now one has to solve the equation (\ref{uF}) in order to find
a function $F(u)$. This function contains all the information
about an analytic structure of an S-matrix. As usually, a solution
of (\ref{uF}) is not unique. Namely, one can multiply it by any
CDD - factor.
In order to remove this ambiguity one should apply a minimality
principle \cite{zamol}.
A combination of this principle with specialities of our problem
gives the following answer
\a
F(\th)={\sinh\o\sinh\L
\over\pi^2}\;Q(\th)\prod_{n=1}^{\infty}{Q(i\pi n +(-1)^n\th)
\over Q(i\pi n -(-1)^n\th)},
\label{fi1}
\b
\a
Q(\th)\;\equiv\;\Gamma [{\L\over\pi^2}(\al-\th)]
\Gamma [1-{\L\over\pi^2}(\al+\th)]\Gamma [{\L\over\pi^2}(\be-\th)]
\Gamma [1-{\L\over\pi^2}(\be+\th)]
\label{fi2}
\b
\a
\al^{(e,h)}_{N,p}\equiv i\pi\o^{(e,h)}_{N,p}/\L^{(e,h)}_{N,p},
\label{fi3}
\b
\a
\be^{(e,h)}_{N,p}\equiv i\pi + \pi ^2/\L^{(e,h)}_{N,p}.
\label{fi4}
\b
The final answer is given by (\ref{sw}), (\ref{crossing}) - (\ref{fi4}).
Let us note that with the only exception
($\M^{(e,h)}_{N,N-1}$) the S-matrix has two poles in the physical
sheet at the points $\al$ and $i\pi-\al$. The first of them
corresponds to a kink bound state in the $s$-channel whose
mass divided by the mass of the fundamental kink
\a
\gamma\;=\;2\cos\al^{(e,h)}_{N,p}/2
\label{ratio}
\b
depends on $p$. This is the main difference of the theories in
external fields from the thermal ones. In the thermal theories
$\gamma$ is stable \cite{gep1}. This observation is connected with
a nonrenormalization property of a mass ratio's
in the $\G^{(1)}$-type ATFT and the opposite property of
the non-simly-laced or twisted ATFT.

\section{Comparison with the known results and further examples}

Among RCFT of $SO(N)$-type there are three well-known models.
The first one - $\M_{N,N-1}$ - lies on the critical line in the
parametric space of Ashkin-Teller model and can be described by
the free massless
scalar field $\phi (z, \bar z)$ compactified on orbifold.
The second one - $\M_{N,N}$ - coincides modulo some irrelevant
subtleties with $Z_{2N}$ parafermionic theory.
Finally, the third one - $\M_{N,\infty}$ - is the model of $N$
free fermions.
It gives us a nice opportunity to check our formulas comparing
them to the results for integrable deformations of these
models.
$${\bf \M^{(e)}_{N,N-1}}$$
The primary field $\Phi^{(e)}_{N,N-1}$ is marginal and coincides with
U(1) current-current operator
$$ \Phi^{(e)}_{N,N-1} = \del \phi \delb \phi.$$
The perturbation does not shift the theory from the critical point.
It should mean that the S-matrix exists only in conformal limit
$\th\to\infty$.
For the crossing parameter we have (\ref{crossing})
$$ \L^{(e)}_{N,N-1} = 0.$$
Therefore rapidity $\th = i\pi u/\L$ becomes infinite as it should
 in agreement with the above observation.
$${\bf \M^{(h)}_{N,N-1}}$$
This model corresponds to $q=e^{i\pi/N-1}$, hence only two
representations - $\pi_0$ and $\pi_{\o}$ - are allowed by the
selection rule (\ref{w}). The admissibility graph (fig.2a))
consists of two points connected by a link. It defines a $Z_2$
 fusion ring
\begin{figure}
\centering
\begin{picture}(200,200)(-100,-120)

\put(19,-80){\makebox(0,0){$\cdots$}}
\put(19,0){\makebox(0,0){$\cdots$}}
\put(-22,85){\makebox(0,0){$\pi_0$}}
\put(18,85){\makebox(0,0){$\pi_{\omega}$}}
\put(-80,33){\makebox(0,0){$\pi_{2\omega}$}}
\put(-80,-33){\makebox(0,0){$\pi_0$}}
\put(-40,5){\makebox(0,0){$\pi_1\equiv\pi_{\omega}$}}
\put(-10,5){\makebox(0,0){$\pi_2\equiv\pi_{\theta}$}}
\put(40,5){\makebox(0,0){$\pi_n\equiv\pi_s$}}
\put(-80,-47){\makebox(0,0){$\pi_{2\omega}$}}
\put(-80,-113){\makebox(0,0){$\pi_0$}}
\put(-40,-75){\makebox(0,0){$\pi_1\equiv\pi_{\omega}$}}
\put(-10,-75){\makebox(0,0){$\pi_2\equiv\pi_{\theta}$}}
\put(40,-75){\makebox(0,0){$\pi_n\equiv\pi_{s+\bar s}$}}
\put(76,-47){\makebox(0,0){$\pi_{2s}$}}
\put(76,-113){\makebox(0,0){$\pi_{2\bar s}$}}
\put(-100,80){\makebox(0,0){$a)$}}
\put(-100,0){\makebox(0,0){$b)$}}
\put(-100,-80){\makebox(0,0){$c)$}}

\put(20,80){\circle*{2}}
\put(-20,80){\circle*{2}}
\put(50,0){\circle*{2}}
\put(-50,0){\circle*{2}}
\put(-10,0){\circle*{2}}
\put(-78,28){\circle*{2}}
\put(-78,-28){\circle*{2}}
\put(50,-80){\circle*{2}}
\put(-50,-80){\circle*{2}}
\put(-10,-80){\circle*{2}}
\put(-78,-52){\circle*{2}}
\put(-78,-108){\circle*{2}}
\put(78,-52){\circle*{2}}
\put(78,-108){\circle*{2}}
\put(60,0){\circle{20}}

\thinlines
\put(-20,80){\line(1,0){40}}
\put(-50,0){\line(-1,1){28}}
\put(-50,0){\line(-1,-1){28}}
\put(-50,0){\line(1,0){40}}
\put(-10,0){\line(1,0){20}}
\put(50,0){\line(-1,0){20}}
\put(-50,-80){\line(-1,1){28}}
\put(-50,-80){\line(-1,-1){28}}
\put(-50,-80){\line(1,0){40}}
\put(-10,-80){\line(1,0){20}}
\put(50,-80){\line(-1,0){20}}
\put(50,-80){\line(1,1){28}}
\put(50,-80){\line(1,-1){28}}

\end{picture}
\caption[x]{\footnotesize Admissibility graphs:

a) $G=SO(N)$, $k=1$

b) $G=SO(2n+1)$, $k=2$

c) $G=SO(2n)$, $k=2$

The b), c) graphs coincide with the McKay correspondence
 graphs of the dihedral groups $d_{4n+2}$, $d_{4n}$.}
\label{fig2}
\end{figure}

\a
\pi_0\otimes\pi_0=\pi_0\0\\
\pi_0\otimes\pi_{\o}=\pi_{\o}\0\\
\pi_{\o}\otimes\pi_{\o}=\pi_0\0\\
\phi_0=\phi_{\o}=1\0
\b
It means that a fundamental
kink is effectively equivalent to a scalar particle \cite{sm1}.
The UV-limit of this particle is given by
$\Phi^{(e)}=\del\phi\delb\phi$.
In other words the fundamental particle S-matrix coincides
with that for the lightest
breather from the SG model at
$$\be^2/8\pi=dim\; \Phi^{(h)}_{N,N-1}=1/2N.$$
The general answer for this case reduces to
\ax
S(\th)=F^{(h)}_{N,N-1}(\th)\;\W{0}{\o}{\o}{0}{\L\th/i\pi}\0\\
\\
\W{0}{\o}{\o}{0}{\L\th/i\pi}=
-{\sinh\Big({\pi i\over N-1} - {2N-1\over 2N-2}\th\Big)
\sinh\Big( {2N-3\over 2N-2}\pi i - {2N-1\over 2N-2}\th\Big)
\over \sin{\pi \over N-1} \sin{\pi \over 2N-2}}
\0
\bx
Analytic structure of the $F$-function is somewhat special. Namely,
besides the ordinary simple poles in the physical sheet at
$\th=\al,i\pi-\al$ there are two additional ones at $\th=\be,i\pi-\be$
, where
\ax
\al^{(h)}_{N,N-1}={2\pi i\over 2N-1}\0\\
\be^{(h)}_{N,N-1}={\pi i\over 2N-1}\0
\bx
At $\al$ - poles the S-matrix reduces to a projector on adjoint
representation of $SO(N)_q$. The cancellation of these poles by
zero's of the $W$ - function agrees with the absence of $\pi_{\th}$
in the fusion ring. So, the only poles are $\be$-poles corresponding
to a scalar bound state.
\a
S^{(h)}_{N,N-1}(\th) = {\sinh\th + i\sin {\pi\over 2N-1}\over
\sinh\th - i\sin {\pi\over 2N-1}},
\b
what coincides with the S-matrix for the lightest SG breather.
$${\bf \M^{(e)}_{N,N}}$$
This theory coincides with $Z_{2N}$ parafermions perturbed by
$$\Phi^{(e)}_{N,N}=
\psi_1\bar{\psi}_1 + \psi^{\dagger}_1\bar{\psi^{\dagger}_1},$$
where $\psi_1(z)$ denotes the first parafermionic current.
Such a theory has been solved in \cite{fat}. This kink theory
has no particle representation, therefore it can be defined
by admissibility graph and corresponding $W$-functions in many
equivalent ways. However, the
analytic structure fixed by $F(\th)$ does not depend on a
representation. In our case the solution is given by (\ref{crossing}
-\ref{fi4})
with
\ax
\L^{(e)}_{N,N} = {i\pi\over 2N+2}\0\\
\o^{(e)}_{N,N} = - {i\pi\over N+1}\0
\bx
The $F$ - function is equal to
$$ F^{(e)}_{N,N}(\th) = {\sin{\pi \over 2N+2}\;\sin{\pi
\over N+1}\over \sinh{i\pi - \th\over 2N+2}\;\sinh{2\pi i
+ \th\over 2N+2}},
$$
what exactly coincides with the unitarizing factor  found in
\cite{fat}.
The admissibility graph for $N=3$ is drawn in fig. 3.

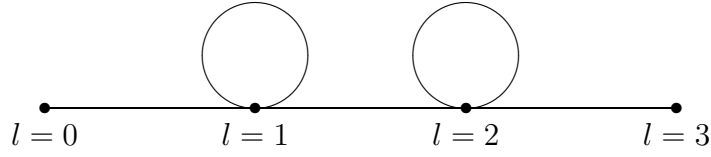
\begin{figure}
\centering
\begin{picture}(200,100)(-100,-50)

\put(-60,-5){\makebox(0,0){$l=0$}}
\put(-20,-5){\makebox(0,0){$l=1$}}
\put(20,-5){\makebox(0,0){$l=2$}}
\put(60,-5){\makebox(0,0){$l=3$}}

\put(-60,0){\circle*{2}}
\put(-20,0){\circle*{2}}
\put(20,0){\circle*{2}}
\put(60,0){\circle*{2}}
\put(20,10){\circle{20}}
\put(-20,10){\circle{20}}

\thinlines
\put(-60,0){\line(1,0){40}}
\put(-20,0){\line(1,0){40}}
\put(20,0){\line(1,0){40}}

\end{picture}
\caption[x]{\footnotesize Admissibility graph for $G=SO(3)$
 at level $k=3$.}
\label{fig3}
\end{figure}

$${\bf \M^{(h)}_{N,N}}$$
This theory coincides with
thermalized parafermions described by the Koberle-Swieca
S-matrix \cite{kobsw}.
It is very interesting to see how this S-matrix
follows from the general answer and we will discuss it in
some more detail. The fusion ring is described by the
admissibility graphs depicted in fig. 2b),c). In notations
given on these pictures one has
\ax
N=2n+1\0\\
\pi_1\otimes \pi_1 = \pi_0 + \pi_{2\o} + \pi_2\0\\
\pi_2\otimes\pi_1 = \pi_1 + \pi_3\0\\
\cdots\0\\
\pi_{n-1}\otimes\pi_1 = \pi_{n-2} + \pi_n\0\\
\pi_{n}\otimes\pi_1 = \pi_{n-1} + \pi_n\0\\
\pi_{2\o}\otimes\pi_1 = \pi_0\otimes\pi_1
=\pi_1\0
\bx
\ax
N=2n\0\\
\pi_1\otimes \pi_1 = \pi_0 + \pi_{2\o} + \pi_2\0\\
\pi_2\otimes\pi_1 = \pi_1 + \pi_3\0\\
\cdots\0\\
\pi_{n-2}\otimes\pi_1 = \pi_{n-3} + \pi_{n-1}\0\\
\pi_{n-1}\otimes\pi_1 = \pi_{n-2} + \pi_{2s} + \pi_{2\bar s}\0\\
\pi_{2\o}\otimes\pi_1 = \pi_0\otimes\pi_1
=\pi_1\0\\
\pi_{2s}\otimes\pi_1 = \pi_{2\bar s}\otimes\pi_1
=\pi_{n-1}\0,
\bx
where $s$ and $\bar s$ denote spinorial and antispinorial
representions. Remarkable fact is that these fusion rings
coincide with the ring of representations of the dihedral
group $d_{2N}$. We will denote them by the same letters.
The dihedral group is the symmetry group of an $N$-gon
and consists of the $2N$ elements
$$ d_{2N} \equiv [\ep_k, \bar{\ep}_k; k=1,\cdots,N].$$
The coincidence we are talking about is achieved provided
\ax
\pi_1(\ep_1) = \left(\brr{cc}e^{2\pi i/N}&0\\
0&e^{-2\pi i/N}\err\right)\0\\
\pi_1(\bar{\ep}_1) = \left(\brr{cc}0&e^{2\pi i/N}\\
e^{-2\pi i/N}&0\err\right)\0
\bx
Dimensions of the representions of $d_{2N}$ coincide with
quantum dimensions of the corresponding representations
of $SO(N)_q$ at level $k=2$. Namely,
\ax
\phi_1=\phi_2=\cdots=2\0\\
\phi_0=\phi_{2\o}=\phi_{2s}=\phi_{2\bar s}=1\0
\bx
It happens something very similar to the situation
described in \cite{sm2}.  The $\l=0$
component of the kink Hilbert
space can be rearranged as a Hilbert space of particles
 and the fundamental kink behaves
like a doublet of scalar particles forming
the $\pi_1$ representation of $d_{2N}$.
The poles of the S-matrix are placed at
\ax
\al^{(h)}_{N,N}=i\pi/N\0
\bx
and the S-matrix describing a scattering of the fundamental
doublets is given by
\ax
S_{11}(\th)=S_{\bar 1 \bar 1}(\th)={\sinh(i\pi/2N + \th/2)\over
\sinh( i\pi/2N-\th/2)}\0\\
S_{1\bar 1}(\th)=S_{\bar 11}(\th)=S_{11}(i\pi-\th)\0
\bx
in agreement with \cite{kobsw}.
$${\bf \M^{(e,h)}_{N,\infty}}$$
In this limit both models - $\M^{(e)}$ and $\M^{(h)}$ -
coincide with $N$ free massive fermions, therefore the S-matrix
should become trivial. Let us demonstrate that it also
follows from the general answer.
The $p\to\infty$ limit means that
\ax
\o\to 0\0\\
\L\to i\pi/2\0\\
a_{\mu}\to 0\0\\
a_{\mu}/\o\to\infty.\0\\
\bx
The last relation is needed to neglect a "boundary effect" of the
admissibility graph. Effectively it should look like an infinite
$N$ - dimensional cubic lattice.
The quantum $SO(N)_q$ - group becomes classical and the kinks
become vector $SO(N)$ - particles.
Local probabilities $W$ diverge in this limit and unitarizing
function $F(\th)$ vanishes in such a ratio that their product
remains finite and diagonal. Namely,
\ax
\Sm{a}{b}{c}{d}{\th} = \delta_{a-c,b-d}\0
\bx
or in the vertex form
\ax
S^{kl}_{ij}(\th) = \delta_{ik}\;\delta_{jl}\0.
\bx
This example concludes the checks of the S-matrix.

Let us pay some attention to $\M^{(e,h)}_{4,p}$ theories
describing an integrable coupling of two mimimal CFT
(\ref{so1}). For example, $\M^{(e)}_{4,5}$ corresponds to
bilayered $Z_3$ Potts models with the energy-energy coupling
between the two layers. The kink - kink scattering for this
physically interesting model immediately follows from
the general answer (Sec. 5). The $G$ - graph is presented
on fig. 4, where we denote $SO(4)$ representation by quantum
numbers of constituting $SU(2)$ representations. The S-matrix
is given by (\ref{sw}),(\ref{crossing})-(\ref{fi4}) with

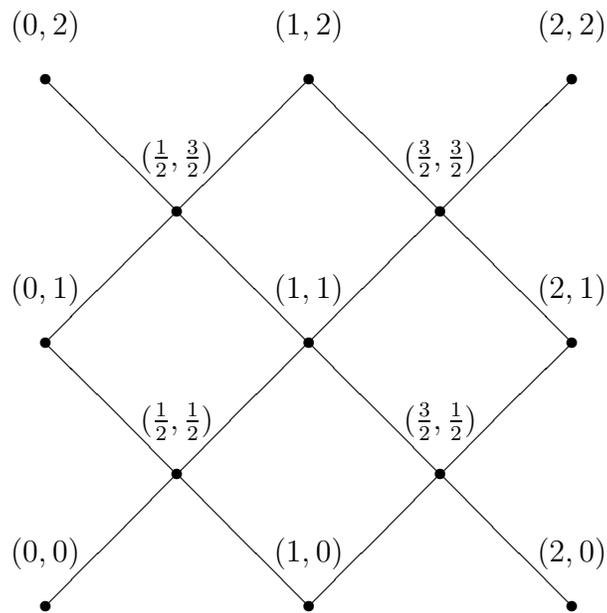
\begin{figure}
\centering
\begin{picture}(200,200)(-100,-100)

\put(-50,10){\makebox(0,0){$(0,1)$}}
\put(0,10){\makebox(0,0){$(1,1)$}}
\put(50,10){\makebox(0,0){$(2,1)$}}
\put(-50,60){\makebox(0,0){$(0,2)$}}
\put(0,60){\makebox(0,0){$(1,2)$}}
\put(50,60){\makebox(0,0){$(2,2)$}}
\put(-50,-40){\makebox(0,0){$(0,0)$}}
\put(0,-40){\makebox(0,0){$(1,0)$}}
\put(50,-40){\makebox(0,0){$(2,0)$}}
\put(-25,35){\makebox(0,0){$({1\over 2},{3\over 2})$}}
\put(25,35){\makebox(0,0){$({3\over 2},{3\over 2})$}}
\put(-25,-15){\makebox(0,0){$({1\over 2},{1\over 2})$}}
\put(25,-15){\makebox(0,0){$({3\over 2},{1\over 2})$}}

\put(-50,0){\circle*{2}}
\put(0,0){\circle*{2}}
\put(50,0){\circle*{2}}
\put(-50,50){\circle*{2}}
\put(0,50){\circle*{2}}
\put(50,50){\circle*{2}}
\put(-50,-50){\circle*{2}}
\put(0,-50){\circle*{2}}
\put(50,-50){\circle*{2}}
\put(25,25){\circle*{2}}
\put(-25,25){\circle*{2}}
\put(25,-25){\circle*{2}}
\put(-25,-25){\circle*{2}}

\thinlines
\put(-50,-50){\line(1,1){100}}
\put(50,-50){\line(-1,1){100}}
\put(0,50){\line(1,-1){50}}
\put(0,50){\line(-1,-1){50}}
\put(0,-50){\line(-1,1){50}}
\put(0,-50){\line(1,1){50}}

\end{picture}
\caption[x]{\footnotesize Admissibility graph for the kink theory
 describig two critical $Z_3$ - models coupled by the
 energy densities. The vacua are marked by  highest weights
of the $SO(4) = SU(2)\times SU(2)$ representations.}
\label{fig4}
\end{figure}

\ax
\o=-i\pi/6\0\\
\L=i\pi/6.
\bx
For this model $\phi_{\o} =3$ and quantum dimensions of all the
allowed representations are integer. This is a hint that the model
can be reconstructed as a particle scattering theory with
fundamental particles forming a triplet. We hope to discuss this
construction in a separate publication.
It should be mentioned that in $\M^{(e)}$ - models $\al$ - pole
leaves a physical sheet. Hence, no bound states appear in these
models and the fundamental kink-kink S-matrix is complete.

\section{Discussion}

The main property of various integrable kink scattering theories is
that in the rational limit they become particle theories described
by some $2D$ integrable QFT's.
These integrable QFT's are generally theories with an explicite symmetry under
some finite group $\G$. Fundamental particles enter the theories by
$\G$ - multiplets and their S-matrix is of GN - type. For the $SU(N)$ - case
such S-matrices have been suggested long ego \cite{kwkt}.
In a sense the RSOS scattering theories could be viewed
as appropriate
restrictions of the GN - type models, or complex ATFT.
The models suggested in the present paper are special, namely,
they are restrictions of the trivially integrable QFT - free
massive fermions. Nevertheless, restricted theories are highly
nontrivial and exhibit a rich analytic structure.
A natural question is whether these models correspond to
a scaling limit of any statistical RSOS theories
described by elliptic local probabilities. We think that those
could be the recently constructed deluted RSOS models \cite{grwa}.

The S-matrix solution of the models $\M^{(e,h)}$ found in Sec. 5
presents an exact result for non-simply-laced complex ATFT at the
discrete set of values of a coupling constant. A mass ratio
of a fundamental kink and it's bound state expressed by
$$\gamma = 2 \cosh \al (p,N)/2$$
depends on the coupling constant parameter $p$.

For example, if $N=2n$ we should obtain a mass ratio of the first and the
second soliton in $B_n^{(1)}$ ATFT described according to the
theorem of Sec.4 by $A_{2n-1,q}^{(2)}$. This ratio
predicted by our formula (\ref{ratio}) is given by
\ax
m_2/m_1 = \gamma = 2 \cos {\pi\over 2n+p},\\
p\equiv {\be^2\over 4\pi^2 - \be^2},
\bx
where $\be$ is a coupling constant of an imaginary ATFT.
This  is in agreement with perturbative calculations
\cite{delgri,mcw}. Of course, other solitonic mass ratios
can be easily calculated along the same lines. By taking
$N=2n+1$ one can obtain a mass spectrum of another set
of non-simply-laced ATFT corresponding to affine superalgebras
 $A^{(4)}(0,2n)$ described by the symmetry $B^{(1)}(0,n)_q$.

It should be noted that $\M^{(t)}$ - theories
associated with $\G^{(1)}$ - like complex ATFT
do not exhibit any coupling dependence of mass ratio's
even if $\G$ is non-simply-laced \cite{gep1}. For such
theories
$$\gamma=2\cosh \al(\G,p)/2=2\cosh i\pi/c_{\G}$$
and does not depend on $p$. It happens due to a contribution
of (para)fermionic loops which are always present in corresponding
ATFT whenever $\G$ is non-simply-laced \cite{fddV}.
In this sense the RSOS - solutions found here are truly non-simply-laced
ATFT solutions with no fermions added
to the action.
\begin{center}
${\bf Acknowledgements}$
\end{center}

I am indebted to G. Sotkov and G. Mussardo for illuminating
discussions and to A. Babichenko for his help in numerical
examinations of the W - characters.



\end{document}